# Structural and mechanical properties of W-Cu compounds characterized by a neural-network-based potential

Jianchuan Liu,[a] Tao Chen,[b] Sheng Mao [c], and Mohan Chen[b, d, e]

[a.] *School of Electrical Engineering and Electronic Information, Xihua University, Chengdu, 610039, P. R. China.*

[b.] *HEDPS, CAPT, College of Engineering and School of Physics, Peking University, Beijing, 100871, P. R. China*

[c.] *Department of Mechanics and Engineering Science, College of Engineering, Peking University, Beijing, 100871, China.*

[d.] *HEDPS, CAPT, College of Engineering, Peking University, Beijing, 100871, China.*

[e.] *AI for Science Institute, Beijing 100080, P. R. China; DP Technology, Beijing 100080, P. R. China.*

**Abstract**

Tungsten-copper (W-Cu) compounds are widely utilized in various industrial fields due to their exceptional mechanical properties. In this study, we have developed a neural-network-based deep potential (DP) model that covers a wide range of temperatures, ranging from 0 to 3,000 K, and pressures, varying from 0 to 10 GPa. This study presents a model trained using density functional theory data for full concentration $Cu_xW_{100-x}$ compounds. Through this model, we systematically investigate the structural and mechanical properties of W-Cu alloys and have the following findings. First, the bulk modulus ($B$) and Young's modulus ($E$) of W-Cu alloys exhibit a linear decline as the Cu content increases, indicating a softening trend in the $Cu_xW_{100-x}$ compounds as the Cu concentration rises. Second, a higher Cu content results in higher critical strain and lower critical stress for these compounds. A brittle-to-ductile transition in the deformation mode predicted is predicted at around 37.5 at. % Cu content. Third, tensile loading tests in the W-Cu gradient structure reveal that Cu-poor region serves as a barrier, hindering shear band propagation while promoting new shear band formation in the Cu-rich region. The above results from the DP model are anticipated to aid in exploring the physical mechanisms underlying the complex phenomena of W-Cu systems and contribute to the advancement of methodologies for materials simulation.

# 1. Introduction

Tungsten (W) and copper (Cu) are widely used industrial materials with several advantages. For example, W possesses high thermal stability, low coefficient of thermal expansion, and the highest melting point (3695 K)[1]. Meanwhile, copper (Cu) owns high electrical performance and excellent thermal conductivity[2]. Over the past few decades, the binary system of W and Cu has attracted much research interest. For instance, W-Cu compounds have been widely used as high-voltage electrical contacts, heat sinks in microelectronics, welding electrodes, etc[3-9]. In addition, W-Cu compounds are regarded promising candidates for the plasma-facing materials in the design of the International Thermonuclear Experimental Reactor (ITER)[10-12], where Cu serves as a heat sink, effectively dissipating considerable amounts of heat during thermonuclear reactions.

Examining the structural (connectivity and morphology of the Cu network), thermodynamic and mechanical properties of W-Cu compounds with varying Cu contents at the atomic scale is essential for evaluating and predicting material performance[13-17]. For instance, the fully connected Cu network can rapidly dissipate heat throughout the entire material due to Cu's high thermal conductivity, enhancing heat dissipation performance and extending the material's service life[16,17]. To optimize W-Cu material performance, it is necessary to prepare W-Cu interpenetrating network materials with high uniformity and densification[18,19]. However, the difference in thermal expansion coefficients of Cu and W results in high thermal residual stresses, making it challenging to connect Cu with W[18]. The strong immiscibility between Cu and W has become an issue for the lifetime of various W-Cu products[20,21]. To address this issue, an effective method is to add interlayer materials with thermal expansion coefficients intermediate between Cu and W[19]. The ideal interlayer material for the Cu-W interface is the W-Cu alloy, which forms a W-Cu gradient structure. This non-uniform composite material combines the various properties of Cu and W, taking advantage of their intrinsic characteristics. It successfully mitigates thermal stresses in the interface region while retaining the high melting point, hardness, and thermal conductivity of W or Cu[18,22-24].

Theoretically, density functional theory (DFT)[25,26] is a powerful quantum-mechanics-based method for studying the structural properties of W-Cu compound systems at the atomic scale[8,9,21,27,28]. For example, Zhang et al. used the generalized-gradient approximation (GGA) exchange-correlational functional to calculate lattice constants, cohesive energies, and formation energies of $Cu_{75}W_{25}$ and $Cu_{50}W_{50}$ alloys, predicting potential metastable states of $Cu_{75}W_{25}$ alloys[29]. In addition, a phase transition from body-centered-cube (BCC) structure to face-centered-cube (FCC) structure was predicted by DFT when Cu content exceeded 70-75%[29-32], which aligns with reported phase diagrams of W-Cu compounds[31,32]. Furthermore, DFT combined with molecular simulations, such as *ab initio* molecular simulations (AIMD), can account for temperature effects when studying the structural, thermodynamic, and mechanical properties of W-Cu compounds at various temperatures. Unfortunately, due to substantial computational costs, most AIMD studies are limited to several hundred atoms and picoseconds, which are inadequate for examining more structural and dynamic properties of W-Cu compounds.

In this regard, the classical molecular dynamics method is attractive as it allows for the study of larger systems for longer simulation time. However, developing empirical potentials for classical molecular dynamics presents several challenges. In 2003, Gong et al. [33] developed an embedding atom method (EAM) potential to predict the glass-forming ability of W-Cu compounds. They further optimized and developed an EAM[29] (Zhang, 2004) potential for characterizing the metastable structure of W-Cu compounds, and the tensile properties of the W/Cu interface were investigated in a subsequent study[34] using MD simulations. In 2019, Wei and Gong et al. [35] proposed an improved EAM (Wei, 2019) potential capable of reproducing mechanical properties and heat formation for various W-Cu compounds. Furthermore, they investigated the effects of Cu on point defects, dislocation loops, and mechanical properties of W using MD simulations[36]. In 2022, Yang and Gong et al. [37] proposed a new EAM (Yang, 2022) potential to describe the structural energy difference between the FCC and BCC structures of W-Cu compounds based on existing EAM potentials. Therefore, it is challenging to develop a potential function that accurately describes all properties of W-Cu compounds. To construct a more general-purposed model for W-Cu compounds, a new method with high precision and low computational costs is necessary.

With the rapid developments of machine learning methods, robust and flexible approaches have emerged to describe atomic interactions with first-principles accuracy. Consequently, neural-network-based interatomic potentials for various materials have been developed via training data from first-principles calculations[38-41]. In particular, the deep potential (DP) model[39] has been widely used to simulate large-scale systems for a long-time scale with the first-principles accuracy.[42-46]. For example, Dai et al.[42] used the DP method to study the structural and elastic properties of high entropy alloys (($Zr_{0.2}Hf_{0.2}Ti_{0.2}Nb_{0.2}Ta_{0.2}$)C). An interatomic DP potential for the Al-Tb alloy was developed by Tang et al.[47] to examine the structure factors of the $Al_{90}Tb_{10}$ metallic liquid. Wang et al.[46] proposed a new descriptor based on the DP method, which accurately describes the three-body interaction of the system and has been successfully applied to the study of mechanical property degradation of W metal.

In this work, we use DP methods to develop an accurate potential function for W-Cu compounds, covering temperatures ranging from 0 to 3,000 K and pressures ranging from 0 to 10 GPa based on density functional theory data for the complete concentration range of $Cu_xW_{100-x}$ compounds. We construct full-concentration W-Cu compounds and thoroughly investigate their microstructure and mechanical properties at various temperatures using the DPMD method. To understand the evolution of material stresses after introducing a gradient layer and improving the properties of the material, we create several W-Cu gradient structures with Cu contents (*x*) ranging from 0 to 100 in increments of $\Delta x$=12.5. Besides, we also systematically analyze the stress and strain along the interface under tensile loading at different temperatures. Our research offers reliable evidence that the generated DP model can be readily applied to study W-Cu compound systems with *ab initio* accuracy while also overcoming the limitations of the simulation time scale associated with AIMD.

## 2. Methods
### 2.1 Density Functional Theory

DFT calculations were performed by the Vienna ab initio simulation package (VASP 5.4.4)[48,49] using the projector augmented wave (PAW) method[50,51] with the Perdew-Burke-Ernzerhof (PBE) exchange-correlational functional[51]. All of the calculations were carried out with a 650 eV cutoff for the plane-wave basis expansion, and the self-consistent electronic loop uses a $10^{-8}$ eV threshold for the total energy. We used a dense gamma-centered *k*-point grid with a 0.16 $Å^{-1}$ spacing between *k*-points. All the AIMD simulations in this work were performed based on the above DFT settings.

## 2.2 Deep Potential Generation

In this work, we constructed BCC and FCC structures with 2 × 2 × 2 supercells for $Cu_xW_{100-x}$ compounds with x = 0, 12.5, 25.0, … , 100. Next, random perturbations were performed on the atomic coordinates by adding values drawn from a uniform distribution in the range of [-0.01, 0.01]. We also changed the cell vectors by a symmetric deformation matrix that is constructed by adding random noise drawn from a uniform distribution in the range of [-0.03, 0.03]. Five steps of AIMD simulations were performed for all the perturbed structures to produce labeled data with energies, forces, and virial tensors calculated from DFT. These labeled data were used to form the initial data sets.

Next, we employed the Deep Potential Generator (DP-GEN)[52] to generate Deep Potential (DP) models[39] for the W-Cu systems. The initial data were trained using the DeePMD-kit package[53]. For the training process, we used three hidden layers of sizes 25, 50, and 100 for the embedding network. In addition, we used three hidden layers with sizes of 240, 240, and 240 for the fitting network. An exponentially decaying learning rate was chosen, ranging from $1.0 \times 10^{-3}$ to $3.5 \times 10^{-8}$. During the optimization process, the prefactor of the energy (force) term in the loss function changed from 0.02 to 1 (1000 to 1). The DP model was trained for $8.0 \times 10^5$ steps with a cutoff radius of 9 Å. Four DP models were generated for each training process, utilizing the same reference dataset but with different initial parameters for the deep neural network.

Finally, we performed MD simulations with the DP model with temperatures ranging from 0 to 3,000 K and pressures ranging from 0 to 10 GPa to explore new configurations using the LAMMPS package[54]. We did not include perturbations in the initial configurations. More details of exploring each iteration are shown in Table S1. In our MD simulations, the structures with model deviation in the range of [0.11, 0.32] eV/Å were labeled as candidate configurations. A maximum of 30 candidate configurations were selected for every $Cu_xW_{100-x}$ compound. These configurations were added to the training set for the next iteration after calculating the energies, forces, and virial tensor using the DFT method. All iterations were done automatically with the DP-GEN package. The dataset update continued until the proportion of candidate configurations was less than 5% and remained almost unaltered for another few iterations.

After the DP-GEN iterations were converged, we trained the collected data for $1.6 \times 10^7$ steps with hybridized multiple descriptors, which consist of both "se_e2_a" and "se_e3" models. Specifically, the "se_e2_a" embedding model takes the distance between two atoms as an input[53,55], while the "se_e3" embedding model adopts the angles between an atom and its two neighboring atoms as input[46]. We used three hidden layers with the size of neurons being 25, 50, and 100 for the "se_e2_a" descriptor and three hidden layers with the size of neurons being 4, 8, and 6 for the "se_e3" descriptor in the embedding network. The cutoff radius was set to 9 Å for the "se_e2_a" descriptors and 4 Å for the "se_e3" descriptors. The root-mean-square errors (RMSEs) of the energies and forces predicted by the finally DP model is shown in Table S2.

## 3. Results

### 3.1 Accuracy of the DP model

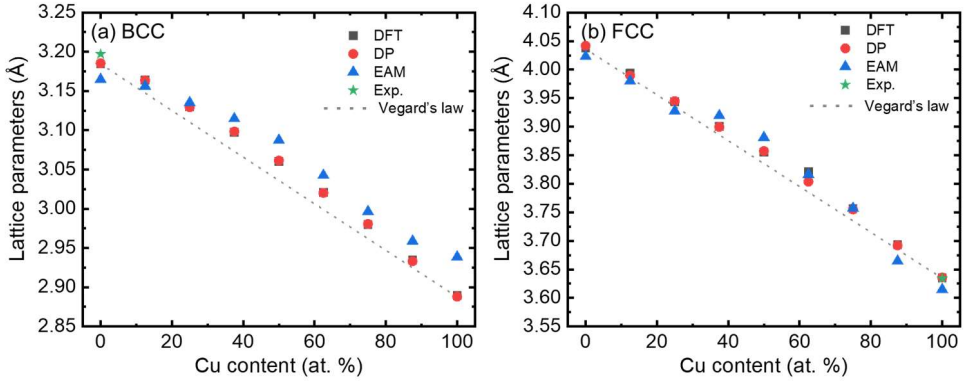

**Fig. 1** Lattice constants of (a) BCC and (b) FCC W-Cu alloys calculated from the DFT, DP, and EAM[37] (Yang, 2022) methods. The experiment values[32] are provided for comparison.

We construct BCC and FCC structures with 2 × 2 × 2 supercells for the $Cu_xW_{100-x}$ compounds with $x$ = 0, 12.5, 25.0, … , 100 by using the special quasi-random structures (SQS) method[56], which is implemented in the Alloy Theoretic Automated Toolkit (ATAT) code[57]. We used three methods to calculate the bulk properties of the W-Cu compounds, i.e., the DFT method, the DP model, and the embedded atom method (EAM) potential. The EAM model parameters are obtained from a recent literature[37] (Yang, 2022).

Fig. 1 shows the calculated lattice parameters of BCC and FCC W-Cu alloys with different concentrations of Cu, as well as available experimental data. As expected, the DP model reproduces the lattice parameters of the BCC and FCC structures, which are in excellent agreement with the DFT results; Both EAM and DP models yield close lattice parameters as compared with the DFT data, while the accuracy of the DP model is slightly better than the EAM model. From Fig. 1, the lattice parameters of BCC and FCC W-Cu alloy are larger than those from Vegard's law[58]. The results obtained by the DP method are consistent with previous works[59], demonstrating that the interactions between W and Cu atoms are mainly repulsive. It also implies that the binary W-Cu alloy is an immiscible system with a large and positive heat of formation[59] (shown in Fig. S1).

**Table 1** Bulk modulus ($B$), shear modulus ($G$), Young's modulus ($E$), and elastic constants ($C_{11}$, $C_{12}$, and $C_{44}$) of $Cu_xW_{100-x}$ as calculated from the DFT, DP, and EAM methods. The errors between DP/EAM and DFT are listed in parentheses.

| Structures | | $B$ (GPa) | $G$ (GPa) | $E$ (GPa) | $C_{11}$ (GPa) | $C_{12}$ (GPa) | $C_{44}$ (GPa) |
|---|---|---|---|---|---|---|---|
| $Cu_{12.5}W_{87.5}$ | DFT | 271 | 188 | 550 | 381 | 216 | 121 |
| (BCC) | DP | 273 (0.6%) | 190 (1.1%) | 555 (0.8%) | 393 (3.1%) | 213 (-1.5%) | 120 (-0.8%) |
| | EAM | 267 (1.6%) | 198 (5.2%) | 553 (0.5%) | 422 (10.7%) | 189 (-12.4%) | 138 (14.3%) |
| $Cu_{25.0}W_{75.0}$ | DFT | 252 | 178 | 514 | 339 | 208 | 129 |
| (BCC) | DP | 259 (2.7%) | 180 (1.3%) | 525 (2.2%) | 334 (-1.6%) | 221 (6.2%) | 132 (2.1%) |
| | EAM | 271 (7.7%) | 194 (8.9%) | 556 (8.2%) | 392 (15.1%) | 211 (1.5%) | 133 (3.2%) |
| $Cu_{37.5}W_{62.5}$ | DFT | 229 | 146 | 452 | 262 | 213 | 90 |
| (BCC) | DP | 227 (-0.7%) | 148 (1.1%) | 452 (-0.1%) | 251 (-4.0%) | 215 (1.2%) | 103 (14.2%) |
| | EAM | 269 (17.2%) | 185 (26.4%) | 544 (20.2%) | 357 (36.2%) | 224 (5.5%) | 125 (38.9%) |
| $Cu_{50.0}W_{50.0}$ | DFT | 199 | 132 | 397 | 268 | 164 | 73 |
| (BCC) | DP | 198 (-0.5%) | 134 (1.8%) | 398 (0.2%) | 242 (-9.7%) | 176 (-7.0%) | 95 (30.4%) |
| | EAM | 183 (-8.0%) | 129 (-1.6%) | 373 (-5.9%) | 199 (-25.5%) | 174 (-6.2%) | 119 (62.5%) |
| $Cu_{62.5}W_{37.5}$ | DFT | 194 | 133 | 393 | 217 | 183 | 108 |
| (BCC) | DP | 196 (0.9%) | 136 (2.2%) | 398 (1.4%) | 223 (2.9%) | 183 (-0.2%) | 113 (4.1%) |
| | EAM | 181 (-6.9%) | 125 (-6.1%) | 366 (-6.7%) | 180 (-16.9%) | 181.43 (-1.1%) | 115 (6.3%) |
| $Cu_{62.5}W_{37.5}$ | DFT | 191 | 126 | 382 | 195 | 189 | 100 |
| (FCC) | DP | 190 (-0.6%) | 126 (0.1%) | 380 (-0.4%) | 212.41 (8.9%) | 179 (-5.6%) | 94 (-6.6%) |
| | EAM | 211 (10.3%) | 145 (14.8%) | 426 (11.8%) | 258.83 (32.7%) | 187 (-1.2%) | 108 (7.7%) |
| $Cu_{75.0}W_{25.0}$ | DFT | 167 | 114 | 337 | 207 | 147 | 82 |
| (FCC) | DP | 169 (-1.3%) | 115 (0.9%) | 341 (1.2%) | 207 (0.1%) | 150.72 (2.2%) | 83 (0.9%) |
| | EAM | 159 (4.9%) | 93 (-17.9%) | 305 (-9.6%) | 173 (-16.2%) | 151 (2.9%) | 40 (-50.9%) |
| $Cu_{87.5}W_{12.5}$ | DFT | 158 | 111 | 321 | 191 | 141 | 89 |
| (FCC) | DP | 156 (-0.7%) | 110 (-0.5%) | 319 (-0.7%) | 191 (0.1%) | 139 (-1.3%) | 89 (-0.6%) |
| | EAM | 156 (-2.2%) | 102 (-7.3%) | 311 (-3.2%) | 191 (-0.2%) | 139 (-1.7%) | 65 (-27.6%) |

Elastic properties are fundamental and indispensable to describe the mechanical properties of materials. They are closely related to essential solid-state properties, such as acoustic velocity and thermal conductivity, etc. Therefore, obtaining accurate elastic properties is crucial to study the mechanical properties of materials. According to previous studies[29,32,37], the BCC and FCC structures of $Cu_xW_{100-x}$ compounds are thermodynamically stable when $0 \leq x \leq 75$ and $75 < x \leq 100$, respectively. Table 1 shows the elastic constants ($C_{11}$, $C_{12}$, and $C_{44}$) of $Cu_xW_{100-x}$ compounds obtained using the DFT, DP, and EAM methods. We find that the elastic constants derived from the DP model are in excellent agreement with the DFT data. However, the elastic constants of $Cu_xW_{100-x}$ compounds obtained by the EAM potential show substantial deviations with the DFT data; note that EAM describes well the elastic constants for pure Cu and pure W[37]. For example, the maximum deviation between DP and DFT results for the elastic constants ($C_{11}$, $C_{12}$, and $C_{44}$) of $Cu_{12.5}W_{87.5}$ is only 3.1% ($C_{11}$), while the deviation between EAM and DFT is 10.7% ($C_{11}$); for $Cu_{87.5}W_{12.5}$, although $C_{11}$ and $C_{12}$ calculated by DP and EAM are close the DFT values, the $C_{44}$ of EAM shows large deviations from the DFT value (-27.6 %). The maximum deviation of $C_{44}$ for EAM is up to 62.5% ($Cu_{50.0}W_{50.0}$). Furthermore, we calculated the bulk modulus ($B$), shear modulus ($G$), and Young's modulus ($E$) using the Voigt-Reuss-Hill approximation [60-62]. From Table 1, we observe that the $B$, $G$, and $E$ calculated by the DP model exhibit excellent agreement with the DFT results (the deviations are within ±3%). However, the deviations from the EAM method are considerably large. For instance, for the shear modulus of $Cu_{37.5}W_{62.5}$ and $Cu_{75.0}W_{25.0}$, the deviations are around 26.4% and -17.9%, respectively. These results demonstrate that the DP model is important for accurately describing the mechanical properties of $Cu_xW_{100-x}$ compounds.

## 4. Discussion
### 4.1 Structural and mechanical properties of $Cu_xW_{100-x}$ compounds

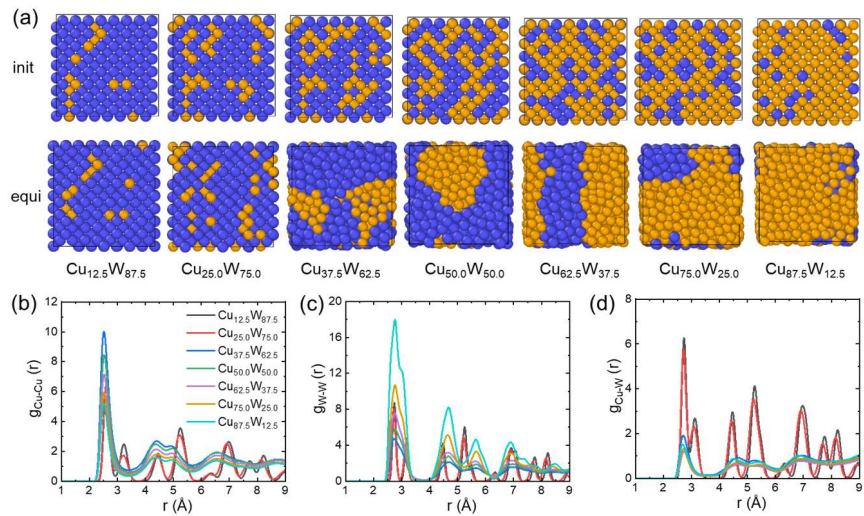

**Fig. 2** (a) Initial (init) and equilibrium (equi) configurations for the $Cu_xW_{100-x}$ compounds containing 1,024 atoms. The blue and yellow balls represent the W and Cu atoms, respectively. (b)-(d) Radial distribution functions g(r) of Cu-Cu, W-W, and W-Cu, as obtained from DPMD simulations of the $Cu_xW_{100-x}$ compounds at 300 K.

Experimentally, W-Cu alloys with lower Cu contents are usually prepared by infiltrating copper into tungsten skeletons[63-65]. However, producing W-Cu alloys containing more than 20% Cu by the infiltration method is challenging[66]. In this regard, the liquid-phase sintering of mixed powders method has been widely employed for fabricating for W-Cu alloys with high Cu contents[67]. We constructed BCC $Cu_xW_{100-x}$ compounds with $x$ = 12.5, 25.0, … , 87.5 containing 1,024 atoms, and the initial structures are shown in the upper row of Fig. 2(a). All of the structures were simulated at 3,000 K and zero external pressure for 0.5 ns with a time step of 5 fs. Next, the structures were quenched to 300 K (a total simulation time of 2.5 ns). Finally, DPMD simulations with a trajectory length of 2.5 ns and a time step of 5 fs were performed, and the lower part of Fig. 2(a) illustrates the final configurations of the $Cu_xW_{100-x}$ compounds. From Fig. 2(a), we observe that the $Cu_xW_{100-x}$ compounds remain as ordered structures when $x \leq 25.0$, but the formation of disordered structure occurs when $x > 25.0$ and the W and Cu domains tend to segregate. The simulation results are consistent with the experimental and calculated results of this compound[9,33,68].

To further examine the local structures of the $Cu_xW_{100-x}$ compounds, Figs. 2(b)-(d) illustrate the radial distribution functions (RDFs) of Cu-Cu, W-W, and W-Cu for different $Cu_xW_{100-x}$ compounds at 300 K, and we have the following findings. First, for the $Cu_{12.5}W_{87.5}$ and $Cu_{25.0}W_{75.0}$ structures with a low concentration of Cu, the RDFs of Cu-Cu, W-W, and W-Cu exhibit a few sharp peaks, suggesting that the configurations remain in the crystalline structures. Second, as the Cu content increases, the $Cu_xW_{100-x}$ compounds form amorphous structures or exhibit phase separation of W and Cu. According to the RDFs of W-W (Fig. 2(c)), we find several peaks at full concentration $Cu_xW_{100-x}$ compounds, indicating that the W domains still maintain a relatively ordered structure. However, the RDF of Cu-Cu (Fig. 2(b)) exhibits only one prominent peak, which indicates that the Cu domains are more disordered as compared to W. Third, we consider the influence of temperature on the structure of the $Cu_xW_{100-x}$ compounds. The details of the changes in the microstructure of the $Cu_xW_{100-x}$ compounds can be observed from the radial distribution functions of the Cu-Cu, W-W, and W-Cu pairs when the temperature rises (Figs. S2-S4). Besides, we find that Cu and W are immiscible even in the liquid state at 2,700 K according to the RDF of W-Cu, and the separation of the Cu and W phases becomes more significant with higher Cu content.

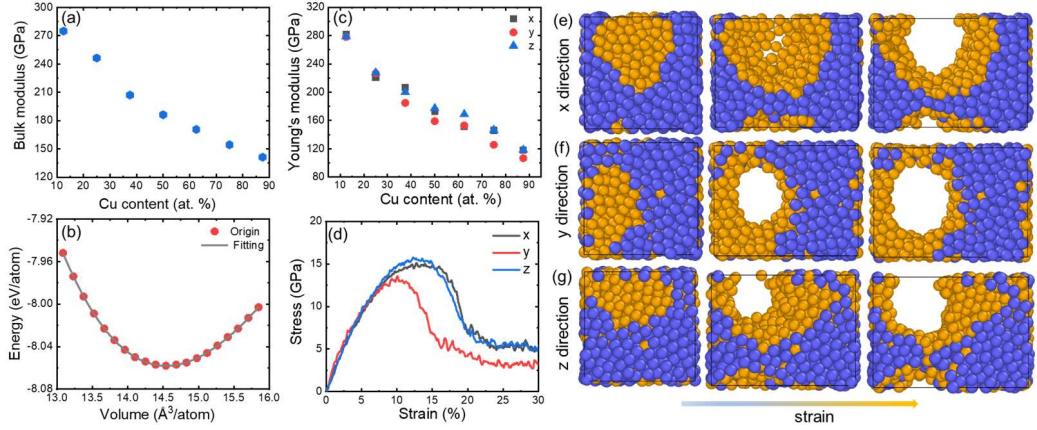

**Fig. 3** (a) Bulk moduli ($B$) of $Cu_xW_{100-x}$ compounds with $x$ ranging from 12.5 to 87.5, and (b) Energy curves *vs* the average atomic volume for the $Cu_{50.0}W_{50.0}$ structure. (c) Young's moduli ($E$) of $Cu_xW_{100-x}$ compounds along the $x$, $y$, and $z$ directions. (d) Strain-stress curves of the $Cu_{50.0}W_{50.0}$ structure under tensile loading along the $x$, $y$, and $z$ directions. (e)-(f) Snapshots of the $Cu_{50.0}W_{50.0}$ structures under tensile loading along the $x$, $y$, and $z$ directions. The blue and yellow balls represent the W and Cu atoms, respectively.

In Table 1, the calculated $B$ and $E$ of the crystalline $Cu_xW_{100-x}$ compounds are listed. Actually, for the $Cu_xW_{100-x}$ compounds with different Cu contents, the structure is mostly amorphous and anisotropic in the $x$, $y$, and $z$ directions. To survey how the Cu contents in a compound affect the elastic properties, we calculate $B$ and $E$ values using the equilibrium configuration after quenching, as shown in Figs. 3(a)(c). The $B$ value is fitting by the curves of energies *vs* the average atomic volume:

$$B = V_0 \left(\frac{\partial^2 E}{\partial V^2}\right), \qquad (1)$$

where $V_0$ depicts the volume of the equilibrium configuration. Fig. 3(b) shows the curves of energies *vs* the average atomic volume for $Cu_{50.0}W_{50.0}$, and the results of other Cu contents show in Fig. S5. The Young's modulus is defined as the slope of the linear portion of a strain-stress curve with the applied strain $\varepsilon < 0.2\%$. Fig. 3(d) shows the strain-stress curves of $Cu_{50.0}W_{50.0}$ under tensile loading along the $x$, $y$, and $z$ directions, and the results of other Cu contents are shown in Fig. S6.

As depicted in Figs. 3(a)(c), both $B$ and $E$ decrease as the Cu contents increase, which indicates a softening trend of the $Cu_xW_{100-x}$ compounds by introducing Cu. The results suggest that the presence of Cu in the W lattice would substantially decrease the tensile strength of the W, which matches well with similar molecular simulation results in the literature[34,37]. Besides, the Young's modulus is about the same in all three directions, i.e., the strain-stress behavior of the $Cu_xW_{100-x}$ compounds exhibits strong isotropy, but the distribution of Cu is anisotropic in the $x$, $y$, and $z$ directions. The results imply that the Young's modulus of a $Cu_xW_{100-x}$ compound is insensitive to the distribution of Cu atoms. From Fig. 3(e), we observe that the Cu domain is necked first, regardless of the loaded tensile direction. In other words, the deformation was mainly concentrated in the Cu domain due to Cu being the soft phase with much lower mechanical strength than W[34]. It indicates that the Cu component has a dramatic effect on the strain-stress behavior of the $Cu_xW_{100-x}$ compounds. According to the strain-stress curve of $Cu_{12.5}W_{87.5}$ and $Cu_{25.0}W_{75.0}$ compounds shown in Fig. S6, the stress drops abruptly upon exceeding the peak, which is attributed to the rapid spontaneous localization of severe plastic shearing into one dominating shear band[69] and the strain energies released. It is shown as a characteristic of poor ductility in $Cu_{12.5}W_{87.5}$ and $Cu_{25.0}W_{75.0}$. As for the comparison of all Cu contents, the strain-stress curve at different Cu contents also reveals that the rate of stress drop tends to be lower with increasing Cu contents, and there exists a brittle-to-ductile transition of the deformation mode at 37.5 at. % Cu contents according to Fig. S6. When the Cu contents exceed 37.5 at. %, the stress reaches a plateau and maintains the same stress level even at $\varepsilon > 20\%$, which indicates an apparent feature of ductile deformation behavior induced by increasing Cu content, i.e., the increasing Cu contents increase the tensile strength of $Cu_xW_{100-x}$ compounds and reduce the brittleness of $Cu_xW_{100-x}$ compounds. As far as we know, the W possesses higher mechanical strength[1] and lower deformation capability than Cu[2]. As a result, increasing of Cu contents in the W-Cu compounds increases the critical strain and decreases the critical stress of $Cu_xW_{100-x}$.

The temperature effects on the tensile properties of $Cu_{50.0}W_{50.0}$ (Fig. S6) show that Young's modulus decreases with the temperature increases. The atomic thermal motion becomes more active with the temperature increase, which overwhelm the binding between atoms[70]. Therefore, the structure of $Cu_{50.0}W_{50.0}$ more easily deforms with temperature increasing, leading to the decrease of Young's modulus. From Fig. S7, we observe that the critical tensile stresses and critical strains decrease with temperature increase. It indicates that the $Cu_xW_{100-x}$ compounds have strong ductility when the temperature increases. But it's important to note that the Cu domain is already in a liquid phase at temperatures above 1,500 K.

## 4.2 Mechanical properties of W-Cu gradient structures

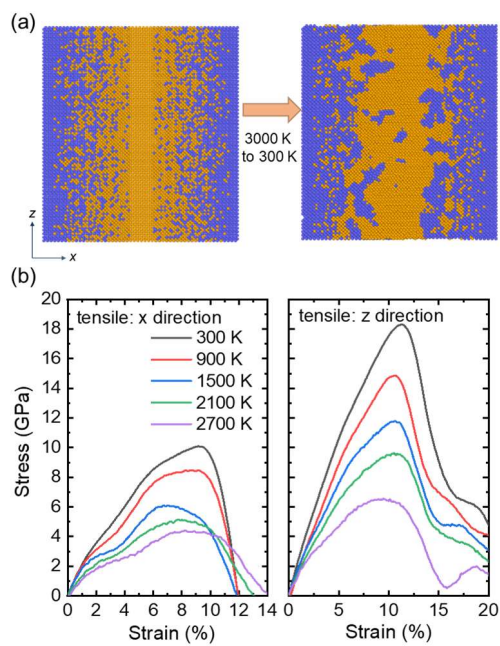

**Fig. 4** (a) Snapshots of the initial and quenched W-Cu gradient structures containing 64,800 atoms. The blue balls represent the W atoms, and the yellow balls depict the Cu atoms. The initial W-Cu gradient structure with the Cu contents $x$ ranging from 0 to 100 with an increase of $\Delta x$=12.5, and a symmetric structure was constructed with the Cu contents $x$ from 100 to 0 decreased by $\Delta x$=12.5 to facilitate the imposition of periodic boundary conditions. (b) Strain-stress curves of the W-Cu gradient structures at different temperatures under tensile loading along the $x$ and $z$ directions.

We constructed a W-Cu gradient structure containing 64,800 atoms, as shown in Fig. 4(a) and Fig. S8. Then, the sample was quenched from 3,000 to 300 K via a 2.5 ns DPMD trajectory using the NVT ensemble. Fig. 4(a) illustrates the cell, which has a size of 165.64×30.67×184.04 Å$^3$ at 300 K. According to the number of densities for Cu and W atoms after quenched (Fig. S9), we find that the simulation cell remains a gradient structure. Fig. 4(b) shows the strain-stress curves of the quenched W-Cu gradient structure under tensile loading along the x and z directions at different temperature ranging from 300 to 2700 K. The results suggest that the critical tensile stresses and critical strains decrease when the temperature rises. In addition, the critical tensile stresses in the *x*-direction are smaller than in the *z*-direction. At 300 K, Young's modulus at *x* direction is 158 GPa, which is close to the experimental value of pure Cu (138 GPa)[71,72], confirming the tensile strength of the W-Cu nano-multilayers material is also mainly dependent on the Cu domain under tensile loading along the *x* direction (the gradient direction).

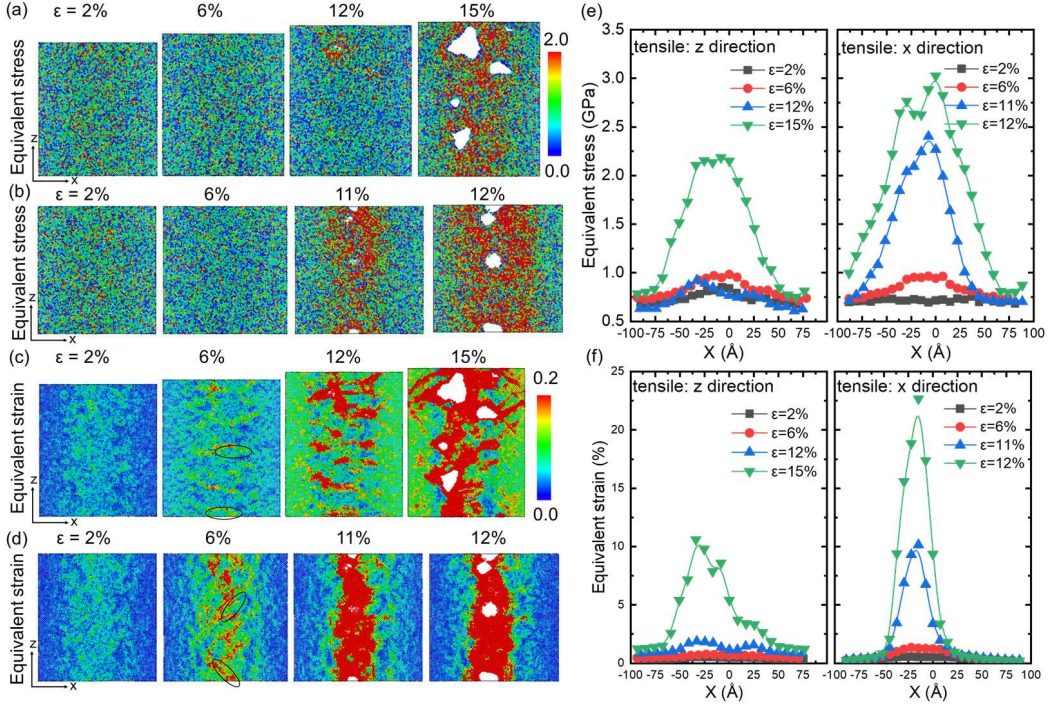

**Fig. 5** Deformation snapshots at 300 K, (a) and (c) for tensile loading along the *z*-axis at the strains of $\varepsilon$ = 2%, 6%, 12%, and 15%, (b) and (d) for tensile loading along the *x*-axis at the overall strains of ε = 2%, 6%, 11%, and 12%. The rainbow nephograms of (a)-(b) and (d)-(c) represent the equivalent stress and equivalent strain for every atom, respectively. (e) Equivalent stress and (f) equivalent strain along the *x*-axis at varied overall strains; the whole gradient structure is divided into several slabs with the same thickness (7.7 Å) in the *x*-axis (the gradient direction), and the averages of the equivalent stress and the equivalent strain were calculated in each slab.

We introduce the equivalent stress $\sigma_e$[73] to further uncover the stress and strain evolution of the W-Cu gradient structure:

$$\sigma_e = \frac{1}{\sqrt{2}}\sqrt{(\sigma_{xx} - \sigma_{yy})^2 + (\sigma_{xx} - \sigma_{zz})^2 + (\sigma_{yy} - \sigma_{zz})^2 + 6(\sigma_{xy}^2 + \sigma_{xz}^2 + \sigma_{yz}^2)}. \qquad (2)$$

In addition, we use the equivalent strain $\varepsilon_e$[73]:

$$\varepsilon_e = \frac{\sqrt{2}}{3}\sqrt{(\varepsilon_{xx}-\varepsilon_{yy})^2+(\varepsilon_{xx}-\varepsilon_{zz})^2+(\varepsilon_{yy}-\varepsilon_{zz})^2+6(\varepsilon_{xy}^2+\varepsilon_{xz}^2+\varepsilon_{yz}^2)} \qquad (3)$$

where $\sigma_{xx}$, $\sigma_{yy}$, $\sigma_{zz}$, $\sigma_{xy}$, $\sigma_{xz}$, $\sigma_{yz}$ are the stress components while $\varepsilon_{xx}$, $\varepsilon_{yy}$, $\varepsilon_{zz}$, $\varepsilon_{xy}$, $\varepsilon_{xz}$, $\varepsilon_{yz}$ are the strain components. The distributions of equivalent stress and strain of the sample at 300 K are shown in Figs. 5(a)-(d), respectively (the OVITO[74] package was used for visualization). The distribution of equivalent stress and strain along the gradient direction increases with increasing strain. Figs. 5(a) and (b) show that the distribution of equivalent stress at the two ends (left and right in the Figs.5(a) and (b)) regions of Cu-poor is uniform, while the equivalent stress increases at the middle region of Cu-rich. Figs. 5(e) and (f) illustrate the distributions of the equivalent stress and the equivalent strain along the x-axis for tensile loading along the x and z directions, respectively.

Figs. 5(c) and (d) show the gradient distribution of equivalent strain along the x-axis. The equivalent strain is larger in the region of Cu-rich since the dislocations activation and movement occur earlier in this region than that in the region of Cu-poor, and then forming the shear bands (marked by the ellipses at ε = 6% in Figs. 5(c) and (d)). We also find that the propagation of these shear bands tends to be impeded in the Cu-rich region, which is about 37.5 at. % Cu for tensile loading along the z direction and about 50.0 at. % Cu for tensile loading along the x direction. Besides, the impeding effect of the Cu-rich region on the shear band propagation is more pronounced when tensile loading is applied along the x direction. With further deformation, more shear bands are involved in plastic deformation in the middle of the sample at ε = 12% (Fig. 5(c)) and at ε = 11% (Fig. 5(d)). The results demonstrate that the Cu-poor region acts as a barrier, which impedes the propagation of shear bands across this region and promotes the formation and propagation of new shear bands in the Cu-rich region. In addition, Fig. 5(f) shows that the equivalent strain becomes large in the middle of the sample with increasing strain, implying the initial plastic deformation was triggered by the region of Cu-rich due to Cu being relatively soft. With additional loading, the Cu-rich region in the middle begins to show stress concentration and deformation.

## 5. Conclusions

In summary, we have systematically examined the fundamental structural and mechanical properties of W-Cu compounds using the DP model with temperatures ranging from 0 to 3,000 K and pressures ranging from 0 to 10 GPa. The DP model accurately reproduced the lattice parameters and elastic constants of the $Cu_xW_{100-x}$ compounds, observing that $x = 25.0$ composition corresponds to an amorphization transition in the $Cu_xW_{100-x}$ compounds, which is in excellent agreement with previous research. Additionally, we have the following findings. First, both the bulk modulus ($B$) and Young's modulus ($E$) of the $Cu_xW_{100-x}$ compounds exhibit an almost linear decrease as Cu content increases, indicating a softening trend in the $Cu_xW_{100-x}$ compounds due to the introduction of Cu. Furthermore, increasing Cu content results in a higher critical strain and lower critical stress for the $Cu_xW_{100-x}$ compounds, with a brittle-to-ductile transition in the deformation mode occurring at 37.5 at. % Cu content. Besides, the tensile strength of the W-Cu gradient structure is primarily dependent on the Cu domain. The Cu-poor region acts as a barrier, which impedes the propagation of shear bands across this region and promotes the formation and propagation of new shear bands in the region of Cu-rich.

## Conflicts of interest

There are no conflicts of interest to declare.

## Acknowledgments

This work of J.L., T.C., S.M. and M.C. is supported by the National Natural Science Foundation of China under Grant No.12122401, No.12074007 and No. 12272005. The numerical simulations were performed on the high-performance computing platform of CAPT and the "Bohrium" cloud computing platform of DP Technology Co., LTD.

## References

[1] P. Chen, G. Luo, Q. Shen, M. Li and L. Zhang, Thermal and electrical properties of W–Cu composite produced by activated sintering, Materials & Design. 46 (2013) 101-105.
[2] L.T. Kong, X.Y. Li, W.S. Lai, J.B. Liu and B.X. Liu, Interfacial reaction of W/Cu examined by an n-body potential through molecular dynamics simulations, Japanese journal of applied physics. 41 (2002) 4503.
[3] H. Rizzo, T. Massalski and M. Nastasi, Metastable crystalline and amorphous structures formed in the Cu-W system by vapor deposition, Metallurgical Transactions A. 24 (1993) 1027-1037.
[4] M. Callisti, M. Karlik and T. Polcar, Bubbles formation in helium ion irradiated Cu/W multilayer nanocomposites: Effects on structure and mechanical properties, Journal of nuclear materials. 473 (2016) 18-27.
[5] Y. Gao, T. Yang, J. Xue, S. Yan, S. Zhou, Y. Wang, D.T. Kwok, P.K. Chu and Y. Zhang, Radiation

tolerance of Cu/W multilayered nanocomposites, Journal of nuclear materials. 413 (2011) 11-15.

[6] D. Lin, J.S. Han, Y.S. Kwon, S. Ha, R. Bollina and S.J. Park, High-temperature compression behavior of W–10 wt.% Cu composite, International Journal of Refractory Metals and Hard Materials. 53 (2015) 87-91.

[7] C. González and R. Iglesias, Energetic analysis of He and monovacancies in Cu/W metallic interfaces, Materials & Design. 91 (2016) 171-179.

[8] F. Moszner, C. Cancellieri, M. Chiodi, S. Yoon, D. Ariosa, J. Janczak-Rusch and L. Jeurgens, Thermal stability of Cu/W nano-multilayers, Acta Materialia. 107 (2016) 345-353.

[9] F. Vüllers and R. Spolenak, From solid solutions to fully phase separated interpenetrating networks in sputter deposited "immiscible" W–Cu thin films, Acta Materialia. 99 (2015) 213-227.

[10] G. Janeschitz, K. Borrass, G. Federici, Y. Igitkhanov, A. Kukushkin, H. Pacher, G. Pacher and M. Sugihara, The ITER divertor concept, Journal of Nuclear Materials. 220 (1995) 73-88.

[11] M. Shimada, A. Costley, G. Federici, K. Ioki, A. Kukushkin, V. Mukhovatov, A. Polevoi and M. Sugihara, Overview of goals and performance of ITER and strategy for plasma–wall interaction investigation, Journal of nuclear materials. 337 (2005) 808-815.

[12] G. Janeschitz and I. Jct, Plasma–wall interaction issues in ITER, Journal of Nuclear Materials. 290 (2001) 1-11.

[13] W. Wang and K. Hwang, The effect of tungsten particle size on the processing and properties of infiltrated W-Cu compacts, Metallurgical and Materials transactions A. 29 (1998) 1509-1516.

[14] M.P. Echlin, A. Mottura, M. Wang, P.J. Mignone, D.P. Riley, G.V. Franks and T.M. Pollock, Three-dimensional characterization of the permeability of W–Cu composites using a new "TriBeam" technique, Acta Materialia. 64 (2014) 307-315. https://doi.org/10.1016/j.actamat.2013.10.043.

[15] J. Fan, T. Liu, S. Zhu and Y. Han, Synthesis of ultrafine/nanocrystalline W–(30–50)Cu composite powders and microstructure characteristics of the sintered alloys, International Journal of Refractory Metals and Hard Materials. 30 (2012) 33-37. https://doi.org/10.1016/j.ijrmhm.2011.06.011.

[16] W. Chen, L. Dong, Z. Zhang and H. Gao, Investigation and analysis of arc ablation on WCu electrical contact materials, Journal of Materials Science: Materials in Electronics. 27 (2016) 5584-5591.

[17] F.A. da Costa, A.G.P. da Silva and U.U. Gomes, The influence of the dispersion technique on the characteristics of the W–Cu powders and on the sintering behavior, Powder Technology. 134 (2003) 123-132.

[18] J. Chapa and I. Reimanis, Modeling of thermal stresses in a graded Cu/W joint, Journal of nuclear materials. 303 (2002) 131-136.

[19] A. Mortensen and S. Suresh, Functionally graded metals and metal-ceramic composites: Part 1 Processing, International materials reviews. 40 (1995) 239-265.

[20] G. Ma, J. Fan and H. Gong, Fundamental effects of hydrogen on cohesion properties of Cu/W interfaces, Solid State Communications. 250 (2017) 79-83.

[21] C.P. Liang, J.L. Fan and H.R. Gong, Cohesion strength and atomic structure of W-Cu graded interfaces, Fusion Engineering and Design. 117 (2017) 20-23. https://doi.org/10.1016/j.fusengdes.2017.02.032.

[22] Y. Itoh, M. Takahashi and H. Takano, Design of tungsten/copper graded composite for high heat flux components, Fusion Engineering and Design. 31 (1996) 279-289.

[23] Y. Lian, X. Liu, Z. Xu, J. Song and Y. Yu, Preparation and properties of CVD-W coated W/Cu FGM

mock-ups, Fusion Engineering and Design. 88 (2013) 1694-1698.

[24] J.-H. You, A. Brendel, S. Nawka, T. Schubert and B. Kieback, Thermal and mechanical properties of infiltrated W/CuCrZr composite materials for functionally graded heat sink application, Journal of nuclear materials. 438 (2013) 1-6.

[25] P. Hohenberg and W. Kohn, Inhomogeneous electron gas, Physical review. 136 (1964) B864.

[26] W. Kohn and L.J. Sham, Self-consistent equations including exchange and correlation effects, Physical review. 140 (1965) A1133.

[27] X.B. Ye, W.Y. Ding, H.Y. He, R. Ding, J.L. Chen and B.C. Pan, An empirical law for the elastic moduli of component-segregated W/Cu compounds, Journal of Alloys and Compounds. 766 (2018) 349-354. https://doi.org/10.1016/j.jallcom.2018.06.345.

[28] N. Selvakumar and S. Vettivel, Thermal, electrical and wear behavior of sintered Cu–W nanocomposite, Materials & Design. 46 (2013) 16-25.

[29] R.F. Zhang, L.T. Kong, H.R. Gong and B.X. Liu, Comparative study of metastable phase formation in the immiscible Cu–W system by ab initio calculation and n-body potential, Journal of Physics: Condensed Matter. 16 (2004) 5251-5258. https://doi.org/10.1088/0953-8984/16/29/016.

[30] C.P. Liang, C.Y. Wu, J.L. Fan and H.R. Gong, Structural, thermodynamic, and mechanical properties of WCu solid solutions, Journal of Physics and Chemistry of Solids. 110 (2017) 401-408. https://doi.org/10.1016/j.jpcs.2017.06.034.

[31] K. Chang, D. Music, M. To Baben, D. Lange, H. Bolvardi and J.M. Schneider, Modeling of metastable phase formation diagrams for sputtered thin films, Sci Technol Adv Mater. 17 (2016) 210-219. https://doi.org/10.1080/14686996.2016.1167572.

[32] K. Chang, M. to Baben, D. Music, D. Lange, H. Bolvardi and J.M. Schneider, Estimation of the activation energy for surface diffusion during metastable phase formation, Acta Materialia. 98 (2015) 135-140. https://doi.org/10.1016/j.actamat.2015.07.029.

[33] H.R. Gong, L.T. Kong, W.S. Lai and B.X. Liu, Glass-forming ability determined by an n-body potential in a highly immiscible Cu-W system through molecular dynamics simulations, Physical Review B. 68 (2003). https://doi.org/10.1103/PhysRevB.68.144201.

[34] G.C. Ma, J.L. Fan and H.R. Gong, Mechanical behavior of Cu-W interface systems upon tensile loading from molecular dynamics simulations, Computational Materials Science. 152 (2018) 165-168. https://doi.org/10.1016/j.commatsci.2018.05.030.

[35] W. Wei, L. Chen, H.R. Gong and J.L. Fan, Strain-stress relationship and dislocation evolution of W-Cu bilayers from a constructed n-body W-Cu potential, J Phys Condens Matter. 31 (2019) 305002. https://doi.org/10.1088/1361-648X/ab1a8a.

[36] W. Wei, C.Y. Wu, J.L. Fan and H.R. Gong, Fundamental effects of copper on dislocation loops and mechanical property of tungsten under irradiation, Journal of Nuclear Materials. 548 (2021) 152838. https://doi.org/10.1016/j.jnucmat.2021.152838.

[37] L. Yang, Y. Shen, S. Mi, J. Fan and H. Gong, Phase stability and mechanical property of W–Cu solid solutions from a newly derived W–Cu potential, Physica B: Condensed Matter. 624 (2022) 413436. https://doi.org/10.1016/j.physb.2021.413436.

[38] Z. Fan, Y. Wang, P. Ying, K. Song, J. Wang, Y. Wang, Z. Zeng, K. Xu, E. Lindgren and J.M. Rahm, GPUMD: A package for constructing accurate machine-learned potentials and performing highly efficient atomistic simulations, arXiv preprint arXiv:2205.10046. (2022).

[39] L. Zhang, J. Han, H. Wang, R. Car and W. E, Deep Potential Molecular Dynamics: A Scalable Model with the Accuracy of Quantum Mechanics, Phys Rev Lett. 120 (2018) 143001.

https://doi.org/10.1103/PhysRevLett.120.143001.

[40] K.T. Schütt, F. Arbabzadah, S. Chmiela, K.R. Müller and A. Tkatchenko, Quantum-chemical insights from deep tensor neural networks, Nat. Commun. 8 (2017) 1-8.

[41] S. Chmiela, A. Tkatchenko, H.E. Sauceda, I. Poltavsky, K.T. Schütt and K.-R. Müller, Machine learning of accurate energy-conserving molecular force fields, Science advances. 3 (2017) e1603015.

[42] F.-Z. Dai, B. Wen, Y. Sun, H. Xiang and Y. Zhou, Theoretical prediction on thermal and mechanical properties of high entropy (Zr0. 2Hf0. 2Ti0. 2Nb0. 2Ta0. 2) C by deep learning potential, Journal of Materials Science & Technology. 43 (2020) 168-174.

[43] W. Liang, G. Lu and J. Yu, Molecular Dynamics Simulations of Molten Magnesium Chloride Using Machine-Learning-Based Deep Potential, Advanced Theory and Simulations. 3 (2020) 2000180. https://doi.org/10.1002/adts.202000180.

[44] J. Huang, L. Zhang, H. Wang, J. Zhao, J. Cheng and W. E, Deep potential generation scheme and simulation protocol for the Li10GeP2S12-type superionic conductors, J Chem Phys. 154 (2021) 094703. https://doi.org/10.1063/5.0041849.

[45] T. Wen, R. Wang, L. Zhu, L. Zhang, H. Wang, D.J. Srolovitz and Z. Wu, Specialising neural network potentials for accurate properties and application to the mechanical response of titanium, npj Computational Materials. 7 (2021) 206. https://doi.org/10.1038/s41524-021-00661-y.

[46] X. Wang, Y. Wang, L. Zhang, F. Dai and H. Wang, A tungsten deep neural-network potential for simulating mechanical property degradation under fusion service environment, Nuclear Fusion. 62 (2022) 126013.

[47] L. Tang, Z.J. Yang, T.Q. Wen, K.M. Ho, M.J. Kramer and C.Z. Wang, Development of interatomic potential for Al-Tb alloys using a deep neural network learning method, Phys Chem Chem Phys. 22 (2020) 18467-18479. https://doi.org/10.1039/d0cp01689f.

[48] G. Kresse and J. Furthmüller, Efficient iterative schemes for ab initio total-energy calculations using a plane-wave basis set, Physical review B. 54 (1996) 11169.

[49] G. Kresse and D. Joubert, From ultrasoft pseudopotentials to the projector augmented-wave method, Physical Review B. 59 (1999) 1758.

[50] P.E. Blöchl, Projector augmented-wave method, Physical review B. 50 (1994) 17953.

[51] J.P. Perdew, K. Burke and M. Ernzerhof, Generalized gradient approximation made simple, Physical review letters. 77 (1996) 3865.

[52] Y. Zhang, H. Wang, W. Chen, J. Zeng, L. Zhang, H. Wang and W. E, DP-GEN: A concurrent learning platform for the generation of reliable deep learning based potential energy models, Computer Physics Communications. 253 (2020) 107206. https://doi.org/10.1016/j.cpc.2020.107206.

[53] H. Wang, L. Zhang, J. Han and W. E, DeePMD-kit: A deep learning package for many-body potential energy representation and molecular dynamics, Computer Physics Communications. 228 (2018) 178-184. https://doi.org/10.1016/j.cpc.2018.03.016.

[54] S. Plimpton, Fast parallel algorithms for short-range molecular dynamics, Journal of Computational Physics. 117 (1995) 1-19.

[55] L. Zhang, J. Han, H. Wang, W.A. Saidi and R. Car, End-to-end symmetry preserving inter-atomic potential energy model for finite and extended systems, arXiv preprint arXiv:1805.09003. (2018).

[56] S.-H. Wei, L. Ferreira, J.E. Bernard and A. Zunger, Electronic properties of random alloys: Special quasirandom structures, Physical Review B. 42 (1990) 9622.

[57] A. Van De Walle, Multicomponent multisublattice alloys, nonconfigurational entropy and other additions to the Alloy Theoretic Automated Toolkit, Calphad. 33 (2009) 266-278.

[58] L. Vegard, Die konstitution der mischkristalle und die raumfüllung der atome, Zeitschrift für Physik. 5 (1921) 17-26.

[59] F. de Boer, R. Boom, W. Mattens and A. Miedema, AK. Niessen, Cohesion in metals, Transition Metal Alloys. (1988).

[60] W. Voigt, Lehrbuch der Kristallphysik, Advances in Earth Science, Advances in Earth Science, 1928, pp.1.

[61] A. Reuß, Berechnung der fließgrenze von mischkristallen auf grund der plastizitätsbedingung für einkristalle, ZAMM-Journal of Applied Mathematics and Mechanics/Zeitschrift für Angewandte Mathematik und Mechanik. 9 (1929) 49-58.

[62] R. Hill, The elastic behaviour of a crystalline aggregate, Proceedings of the Physical Society. Section A. 65 (1952) 349.

[63] P. Chen, G. Luo, M. Li, Q. Shen and L. Zhang, Effects of Zn additions on the solid-state sintering of W–Cu composites, Materials & Design (1980-2015). 36 (2012) 108-112.

[64] A.G. Hamidi, H. Arabi and S. Rastegari, Tungsten–copper composite production by activated sintering and infiltration, International Journal of Refractory Metals and Hard Materials. 29 (2011) 538-541.

[65] E. Uhlmann, S. Piltz and K. Schauer, Micro milling of sintered tungsten–copper composite materials, Journal of Materials Processing Technology. 167 (2005) 402-407.

[66] F. Jinglian, Y. De-jian and H. Bo-yun, Current status of R&D of W-Cu composite materials in China and abroad, Powder Metall. Ind. 13 (2003) 9-14.

[67] J. Johnson and R. German, Phase equilibria effects on the enhanced liquid phase sintering of tungsten-copper, Metallurgical transactions A. 24 (1993) 2369-2377.

[68] S.N. Alam, Synthesis and characterization of W–Cu nanocomposites developed by mechanical alloying, Materials Science and Engineering: A. 433 (2006) 161-168. https://doi.org/10.1016/j.msea.2006.06.049.

[69] L. Zhao, K.C. Chan, S.H. Chen, S.D. Feng, D.X. Han and G. Wang, Tunable tensile ductility of metallic glasses with partially rejuvenated amorphous structures, Acta Materialia. 169 (2019) 122-134. https://doi.org/10.1016/j.actamat.2019.03.007.

[70] J. Liu, Molecular dynamic study of temperature dependence of mechanical properties and plastic inception of CoCrCuFeNi high-entropy alloy, Physics Letters A. 384 (2020) 126516. https://doi.org/10.1016/j.physleta.2020.126516.

[71] Y. Chang and L. Himmel, Temperature dependence of the elastic constants of Cu, Ag, and Au above room temperature, Journal of Applied Physics. 37 (1966) 3567-3572.

[72] T. Hehenkamp, W. Berger, J.-E. Kluin, C. Lüdecke and J. Wolff, Equilibrium vacancy concentrations in copper investigated with the absolute technique, Physical Review B. 45 (1992) 1998.

[73] R. Von Mises, Mechanik der festen Körper im plastisch- deformablen Zustand, Nachr. Ges. Wiss. Gottingen. (1913) 582.

[74] A. Stukowski, Visualization and analysis of atomistic simulation data with OVITO–the Open Visualization Tool, Modelling and simulation in materials science and engineering. 18 (2009) 015012.